\documentclass[twocolumn,showpacs,prl,10pt]{revtex4}
\usepackage[english]{babel}
\usepackage{amsmath,amssymb}
\usepackage{times}
\usepackage{epsfig}

\begin{document}

\title {Spin-fluctuation mechanism of superconductivity in cuprates}
\author{P. Prelov\v sek$^{1,2}$ and A. Ram\v sak$^{1,2}$}
\affiliation{$^1$Faculty of Mathematics and Physics, University
of Ljubljana, SI-1000 Ljubljana, Slovenia}
\affiliation{$^2$J.\ Stefan Institute, SI-1000 Ljubljana,
Slovenia}
\date{January 30, 2005}

\begin{abstract}
The theory of superconductivity within the $t$-$J$ model, as relevant
for cuprates, is developed. It is based on the equations of motion for
projected fermionic operators and the mode-coupling approximation for
the self-energy matrix. The dynamical spin susceptibility at various
doping is considered as an input, extracted from experiments. The
analysis shows that the superconductivity onset is dominated by the
spin-fluctuation contribution. We show that $T_c$ is limited by the
spin-fluctuation scale $\Gamma$ and shows a pronounced dependence on
the next-nearest-neighbor hopping $t'$. The latter can offer an
explanation for the variation of $T_c$ among different families of
cuprates.

\end{abstract}
\maketitle 

Since the discovery of high-temperature superconductivity (SC) in
cuprates the mechanism of SC in these compounds represents one of the
central open questions in the solid state theory. The role of strong
correlations and the antiferromagnetic (AFM) state of the reference
insulating undoped compound has been recognized very early
\cite{ande}. Still, up to date there is no general consensus whether
ingredients as embodied within the prototype single-band models of
strongly correlated electrons are sufficient to explain the onset of
high $T_c$, or in addition other degrees of freedom, as e.g. phonons,
should be invoked. As the basis of our study, we assume the simplest
$t$-$J$ model \cite{bask}, allowing besides the nearest-neighbor (NN)
hopping $t$ also for the next-nearest-neighbor (NNN) hopping $t'$
term. The latter model, as well as the Hubbard model \cite{bulu},
in the strong correlation limit $U \gg t$ closely related to the
$t$-$J$ model, have been considered by numerous authors to address the
existence of SC due to strong correlations alone. Within the parent
resonating-valence-bond (RVB) theory \cite{ande,bask} and slave-boson
approaches to the $t$-$J$ model \cite{kotl} the SC emerges due to the
condensation of singlet pairs, induced by the exchange interaction
$J$.  An alternative view on strong correlations has been that AFM
spin fluctuations, becoming particularly longer-ranged and soft at low
hole doping, represent the relevant low-energy bosonic excitations
mediating the attractive interaction between quasiparticles (QP) and
induce the $d$-wave SC pairing. The latter scenario has been mainly
followed in the planar Hubbard model \cite{bulu} and in the
phenomenological spin-fermion model \cite{mont}.  Recent numerical
studies of the planar $t$-$J$ model using the variational quantum
Monte Carlo approach \cite{sore}, as well as of the Hubbard model
using cluster dynamical mean-field approximation \cite{maie}, seem to
confirm the stability of the $d$-wave SC as the ground state at
intermediate hole doping.

The $t$-$J$ model is nonperturbative by construction, so it is hard to
design for it a trustworthy analytical method. One approach is to use
the method of equations of motion (EQM) to derive an effective
coupling between fermionic QP and spin fluctuations \cite{prel}. The
latter method has been employed to evaluate the self energy and
anomalous properties of the spectral function \cite{prel,plak,prel1},
in particular the appearance of the pseudogap and the effective
truncation of the Fermi surface (FS) at low hole doping
\cite{prel1}. The analysis has been extended to the study of the SC
pairing \cite{plak}, while an analogous approach has been also applied
to the Hubbard model \cite{onod}.

In the following we adopt the formalism of the EQM and the resulting
Eliashberg equations within the simplest mode-coupling approximation
\cite{plak,prel1}. Equations involve the dynamical spin susceptibility
which we consider as an input taken from the
inelastic-neutron-scattering (INS) and NMR-relaxation experiments in
cuprates.  The analysis of these experiments \cite{bonc} reveals that
in the metallic state the AFM staggered susceptibility is strongly
enhanced at the crossover from the overdoped (OD) regime to optimum
(OP) doping and is increasing further in underdoped (UD) cuprates,
while at the same time the corresponding spin-fluctuation energy scale
is becoming very soft. Direct evidence for the latter is the
appearance of the resonant magnetic mode \cite{ross,fong} within the
SC phase indicating that the AFM paramagnon mode can become even lower
than the SC gap.  These facts give a renewed support to the scenario
that the spin fluctuations in cuprates represent the lowest bosonic
mode relevant for the mechanism of the $d$-wave SC pairing. One of the
novel results of our approach is the importance of the NNN hopping
$t^\prime$, consistent with the emerging evidence for the variation of
$T_c$ among different families of cuprates \cite{pava}.

We consider the $t$-$J$ model
\begin{equation}
H=-\sum_{i,j,s}t_{ij} \tilde{c}^\dagger_{js}\tilde{c}_{is}
+J\sum_{\langle ij\rangle}({\bf S}_i\cdot {\bf S}_j-\frac{1}{4}
n_in_j), \label{tj} 
\end{equation}
including both the NN hopping $t_{ij}=t$ and the NNN hopping
$t_{ij}=t^\prime$. The projection in fermionic operators,
$\tilde{c}_{is}= (1-n_{i,-s}) c_{is}$ leads to a nontrivial EQM, which
can be in the ${\bf k}$ basis represented as
\begin{eqnarray}
\label{eqm}
&&[\tilde c_{{\bf k} s},H]= [(1+c_h) \frac{\epsilon^0_{\bf k}}{2} - J
(1-c_h)] \tilde c_{{\bf k} s} + \\ &&\frac{1}{\sqrt{N}}\sum_{\bf q}
m_{\bf k q} \bigl[ s S^z_{\bf q} \tilde c_{{\bf k}-{\bf q},s} +
S^{\mp}_{\bf q} \tilde c_{{\bf k}-{\bf q},-s} -
\frac{1}{2} \tilde n_{\bf q} \tilde c_{{\bf k}-{\bf q}, s}\bigr],
\nonumber
\end{eqnarray}
where $c_h$ is the hole concentration and $m_{\bf k q}$ is the
effective spin-fermion coupling $ m_{\bf k q}=2J \gamma_{\bf q} +
\epsilon^0_{{\bf k}-{\bf q}}$, while $\epsilon^0_{\bf
k}=-4t\gamma_{\bf k}- 4t'\gamma'_{\bf k}$ is the bare band dispersion
on a square lattice.  As argued in Ref.~\cite{prel1} we use the
symmetrized coupling in order to keep a similarity with the
spin-fermion phenomenology \cite{mont}
\begin{equation}
\label{mkq}
\tilde m_{\bf kq}= 2J \gamma_{\bf q}+
\frac{1}{2} (\epsilon^0_{{\bf k}-{\bf q}}+\epsilon^0_{\bf k}).
\end{equation}
EQM, Eq.~(\ref{eqm}), are used to derive the approximation for the
Green's function (GF) matrix $G_{{\bf k}s}(\omega)= \langle\langle
\Psi_{{\bf k}s}| \Psi^\dagger_{{\bf k}s} \rangle\rangle_\omega$ for
the spinor $\Psi_{{\bf k}s}=(\tilde c_{{\bf k},s},\tilde
c^\dagger_{-{\bf k},-s})$.  We follow the method, as applied to the
normal state (NS) GF by present authors \cite{prel,prel1}, and
generalized to the SC pairing by Plakida and Oudovenko \cite{plak}. In
general, we can represent the GF matrix in the form
\begin{equation}
G_{{\bf k}s}(\omega)^{-1}=\frac{1}{\alpha} [\omega \tau_0 -\hat
\zeta_{{\bf k}s} +\mu \tau_3 -\Sigma_{{\bf k}s}(\omega) ], \label{gf}
\end{equation}
where $\alpha = \sum_i \langle \{\tilde c_{i s},\tilde c^{\dagger}_{i
s}\}_+ \rangle/N = (1+c_h)/2$ is the normalization factor, $\mu$ is
the chemical potential and $\hat \zeta_{{\bf k}s}$ the frequency
matrix,
\begin{equation}
\hat \zeta_{{\bf k}s}=\frac{1}{\alpha}\langle \{[\Psi_{{\bf k}s},H],
\Psi^{\dagger}_{{\bf k}s} \}_+ \rangle. \label{zeta}
\end{equation}
Eq.~(\ref{zeta}) generates a renormalized band $\tilde \zeta_{\bf k}=
\zeta^{11}_{{\bf k}s}= \bar \zeta - 4 \eta_1 t \gamma_{\bf k}-
4 \eta_2 t' \gamma'_{\bf k}$ and the mean-field (MF) SC gap
\begin{equation}
\Delta^0_{\bf k}=\zeta^{12}_{{\bf k}s}= - \frac{4J}{N\alpha} 
\sum_{\bf q} \gamma_{{\bf k} -{\bf q}} \langle 
\tilde c_{-{\bf q},-s} \tilde c_{{\bf q},s} \rangle.
\label{del0}
\end{equation}

To evaluate $\Sigma_{{\bf k}s}(\omega)$ we use the lowest-order
mode-coupling approximation, analogous to the treatment of the SC in
the spin-fermion model \cite{mont}, introduced in the $t$-$J$ model
for the NS GF \cite{prel,prel1} and extended to the analysis of the SC
state \cite{plak}. Taking into account EQM, Eq.~(\ref{eqm}), and by
decoupling fermionic and bosonic degrees of freedom, one gets
\begin{eqnarray}
&&\Sigma^{11(12)}_{{\bf k}s}(i \omega_n)=\frac{-1}{N\alpha\beta}
\sum_{\bf q} \sum_m \tilde m^2_{\bf kq} G^{11(12)}_{{\bf k}-{\bf
q},s}(i \omega_m) \times \nonumber \\ &&\qquad [3\chi_{\bf q}(i
\omega_n-i \omega_m) \pm \frac{1}{4} \chi^c_{\bf q}(i \omega_n-i
\omega_m)],
\label{sig}
\end{eqnarray}
where $i \omega_n=i \pi(2n+1)/\beta$ and $\chi_{\bf
q}(\omega),\chi^c_{\bf q}(\omega)$ are dynamical spin and charge
susceptibilities, respectively. In cuprates charge fluctuations seem
to be much less pronounced, so we furtheron keep only the spin
contribution.

In order to analyze the low-energy behavior in the NS and in the SC
state, we use the QP approximation for the spectral function matrix
\begin{equation}
A_{{\bf k}s}(\omega) \sim \frac{\alpha Z_{\bf k}} {2 E_{\bf k}}
(\omega \tau_0 -\epsilon_{\bf k}\tau_3 -
\Delta_{{\bf k}s}\tau_1) [\delta(\omega-E_{\bf k} )-
\delta(\omega+E_{\bf k})], \label{ak}
\end{equation}
where $E_{\bf k}=(\epsilon^2_{\bf k}+\Delta^2_{{\bf k}s})^{1/2}$,
while NS parameters, i.e., the QP weight $Z_{\bf k}$ and the QP
energy $\epsilon_{\bf k}$, are determined from $G_{{\bf k}s}(\omega
\sim 0)$, Eq.~(\ref{gf}).  The renormalized SC gap is
\begin{equation}
\Delta_{{\bf k}s}= Z_{\bf k}[\Delta^0_{\bf k}+\Sigma^{12}_{{\bf k}s}(0)].
\end{equation}
It follows from Eq.~(\ref{gf}) that $G^{12}_{{\bf k}s}(i\omega_n) \sim
- \alpha Z_{\bf k} \Delta_{{\bf k}s}/(\omega_n^2+E^2_{\bf k})$.  By
defining the normalized frequency dependence $F_{\bf
q}(i\omega_l)=\chi_{\bf q}(i\omega_l)/\chi^0_{\bf q}$, and rewriting
the MF gap, Eq.~(\ref{del0}), in terms of the spectral function,
Eq.~(\ref{ak}), we can display the gap equation in a more familiar
form,
\begin{eqnarray}
\Delta_{{\bf k}s}&=&\frac{1}{N}\sum_{\bf q}[4J \gamma_{{\bf k}-{\bf
q}} -3\tilde m^2_{{\bf k},{\bf k}-{\bf q}}\chi^0_{{\bf k}-{\bf q}}
C_{{\bf q},{\bf k}-{\bf q}}] \times \nonumber \\ &&(Z^0_{\bf k}
Z^0_{\bf q} \Delta_{{\bf q}s} /2 E_{\bf q} ) \mathrm{th}(\beta E_{\bf
q}/2), \label{del}
\end{eqnarray}
where $C_{\bf k q}=I_{\bf kq}(i\omega_n \sim 0)/I^0_{\bf k}$ plays the
role of the cutoff function with
\begin{equation}
I_{\bf kq}(i\omega_n)=\frac{1}{\beta} \sum_{m} F_{\bf q}(i\omega_n-
i\omega_m) \frac{1}{\omega_m^2 + E^2_{{\bf k}s}}, \label{ikq}
\end{equation}
and $I^0_{\bf k}=\mathrm{th}(\beta E_{\bf k}/2)/(2 E_{\bf k})$.
Eq.~(\ref{del}) represents the BCS-like expression which we use
furtheron to evaluate $T_c$, as well to discuss the SC gap
$\Delta_{\bf q}(T=0)$. To proceed we need the input of two kinds: a)
the dynamical spin susceptibility $\chi_{\bf q}(\omega)$, and b) the
NS QP properties $Z_{\bf k},\epsilon_{\bf k}$.

The INS experiments show that within the NS the low-$\omega$ spin
dynamics at ${\bf q} \sim {\bf Q}$ is generally overdamped in the
whole doping (but paramagnetic) regime \cite{fong}. Hence we assume
$\chi_{\bf q}(\omega)$ of the form
\begin{equation}
\chi^{\prime\prime}_{\bf q}(\omega)= \frac{B_{\bf q} \omega}
{\omega^2+\Gamma^2_{\bf q}}, \qquad
F_{\bf q}(i\omega_l)= \frac{\Gamma_{\bf q}}{|\omega_l|+
\Gamma_{\bf q} }. \label{chi}
\end{equation}
Following the recent memory-function analysis \cite{sega} $B_{\bf
q}=\chi^0_{\bf q}\Gamma_{\bf q}$ should be quite independent of
$\tilde {\bf q}={\bf q}-{\bf Q}$. We choose the variation as
$\Gamma_{\bf q} \sim \Gamma_{\bf Q}(1+ w \tilde q^2/\kappa^2)^2$
consistent with the INS observation of faster than Lorentzian fall-off
of $\chi^{\prime\prime}_{\bf q}(\omega)$ vs. $\tilde q$
\cite{fong}. $w \sim 0.42$ in order that $\kappa$ represents the usual
inverse AFM correlation length.

Consequently, we end up with parameters $\chi^0_{\bf Q},\Gamma_{\bf
Q},\kappa$, which are dependent on $c_h$, but in general as well vary
with $T$. Although one can attempt to calculate them using the
analogous framework \cite{sega}, we use here the experimental input
for cuprates. We refer to results of the recent analysis \cite{bonc},
where NMR $T_{2G}$ relaxation and INS data were used to extract
$\kappa$, $\chi^0_{\bf Q}(T)$ and $\Gamma_{\bf Q}(T)$ for various
cuprates, ranging from the UD to the OD regime. For comparison with
the $t$-$J$ model, we use usual parameters $t=400$~meV $\sim 4500$~K,
$J=0.3 t$. At least for UD cuprates, quite consistent estimates for
$\chi^0_{\bf Q},\Gamma_{\bf Q}$ can be obtained also directly from the
INS spectra \cite{fong}. For UD, OP and OD regime, i.e., $c_h=0.12,
0.17, 0.22$, respectively, we use furtheron the following values:
$\chi^0_{\bf Q}t=15.0, 4.0, 1.0$, $\Gamma^0_{\bf Q}/t=0.03,0.1,0.18$
(appropriate at low $T$), and $\kappa=0.5,1.0,1.2$. It
is evident, that in the UD regime the energy scale $\Gamma^0_{\bf Q}$
becomes very small (and consequently $\chi^0_{\bf Q} \propto
1/\Gamma^0_{\bf Q}$ large, in spite of modest $\kappa$ \cite{bonc}),
supported by a pronounced resonance mode \cite{fong}. We take into
account also the $T$ dependence, i.e., $\Gamma_{\bf Q}(T) \sim
\Gamma^0_{\bf Q} + T$ \cite{bonc}, being significant only in the UD
regime.

For the NS $A_{\bf k}(\omega)$ and corresponding $Z_{\bf
k},\epsilon_{\bf k}$ we solve Eq.~(\ref{sig}) for $\Sigma^{11}_{\bf
k}=\Sigma_{\bf k}$ as in Ref.~\cite{prel1}, with the input for
$\chi_{\bf q}(\omega)$ as described above. Since our present aim is on
the mechanism of the SC, we do not perform the full self-consistent
calculation of $\Sigma_{\bf k}(\omega)$, but rather simplify it as
done in the previous study \cite{prel1}.  As evident from EQM,
Eq.~(\ref{eqm}), the effective spin-fermion coupling $\tilde m_{\bf
kq}$ is strong (of the order of $t$) generating a strong incoherent
component of the spectral function, leading to an overall decrease of
the QP weight $\bar Z<1$ and the QP dispersion with renormalized
$\eta_1,\eta_2<1$. Soft AFM fluctuations with ${\bf q} \sim {\bf Q}$
lead through Eq.~(\ref{sig}) to an additional reduction of $Z_{\bf
k}$, which is ${\bf k}$-dependent. A pseudogap appears along the AFM
zone boundary and the FS is effectively truncated in the UD regime
with $Z_{{\bf k}_F} \ll 1$ near the saddle points $(\pi, 0)$ (in the
antinodal part of the FS) \cite{prel1}. We fix $\mu$ with the FS
volume corresponding to band filling $1-c_h$.

We first comment general properties of the gap equation,
Eq.~(\ref{del}). Close to half-filling and for $\chi^0_{\bf q}$ peaked
at ${\bf q}\sim {\bf Q}$ both terms favor the $d_{x^2-y^2}$ SC. The
MF-part $\Delta^0_{\bf k}$, Eq.~(\ref{del0}), involves only $J$ which
induces a nonretarded local attraction, playing the major role in the
RVB theories \cite{ande,bask}. In contrast, the spin-fluctuation part
represents a retarded interaction due to the cutoff function $C_{{\bf
k}{\bf q}}$ determined by $\Gamma_{{\bf k}-{\bf q}}$. The largest
contribution to the SC pairing naturally arises from the antinodal
part of the FS. Meanwhile, in the same region of the FS also $Z_{\bf
k}$ is smallest, reducing the pairing strength in particular in the UD
regime.  Our analysis is also based on the lowest order mode-coupling
treatment of the SC pairing as well as of the QP properties near the
FS. Taking this into account, one can question the relative role of
the hopping parameters $t,t^\prime$ and the exchange $J$ in the
coupling, Eq.~(\ref{mkq}). While our derivation within the $t$-$J$
model is straightforward, an analogous analysis within the Hubbard
model using the projections to the lower and the upper Hubbard band,
respectively, would not yield the $J$ term within the lowest order
since $J \propto t^2$. This stimulates us to investigate in the
following also separately the role of $J$ term in Eq.~(\ref{del}),
both through the MF term, Eq.~(\ref{del0}), and the coupling $\tilde
m_{\bf kq}$, Eq.~(\ref{mkq}).

Let us turn to results for the NS spectral properties and consequently
$T_c$. As discussed above we assume that the effective band is
renormalized due to the incoherent hopping \cite{prel1}, with:
$\eta_1=\eta_2=0.5,\bar Z=0.7$, whereas the coupling to low-energy AFM
fluctuations, Eq.~(\ref{sig}), leads to an additional QP
renormalization. For fixed $t'/t=-0.3$, we present in Fig.~1 results
for the variation of the $Z_{\bf k}$ in the Brillouin zone for two
sets of parameters, representing the UD and the OD regime,
respectively. The location of the renormalized FS is also presented in
Fig.~1. While the coupling to AFM fluctuations partly changes the
shape of the FS, more pronounced effect is on the QP weight. It is
evident from Fig.~1 that $Z_{\bf k}$ is reduced along the AFM zone
boundary away from the nodal points. Particularly strong
renormalization $Z_{\bf k} \ll 1 $ happens in the UD case, leading to
an effective truncation of the FS away from nodal points \cite{prel1}.

\begin{figure}[htb]
\centering
\epsfig{file=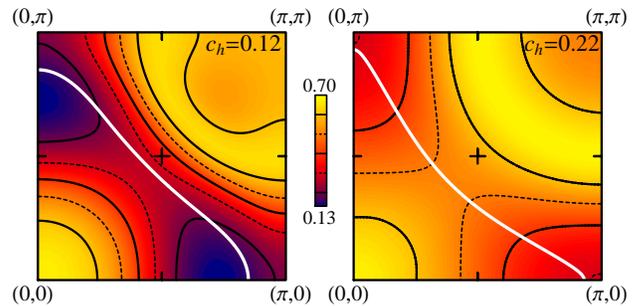,width=90mm,angle=0}
\caption{QP weight $Z_{\bf k}$ evaluated for $t'/t=-0.3$ 
for parameters corresponding to $c_h=0.12$ and $c_h=0.22$,
respectively. White line represents the location of the FS.}
\label{fig1}
\end{figure}

NS results for $Z_{\bf k},\epsilon_{\bf k}$ are used as an input for
the solution of the gap equation, Eq.~(\ref{del}), as presented in
Fig.~2. For the same $t^\prime/t=-0.3$ we calculate $T_c/t$ for
$c_h=0.12,0.17,0.22$. Besides the result a) of Eq.~(\ref{del}) (full
line in Fig.~2) we present also two alternatives: b) the solution of
Eq.~(\ref{del}) without the MF term, and c) the result with $\tilde
m_{\bf kq}$ without the $J$ term and omitted MF term. In the latter
case, we used as input NS QP parameters, recalculated with
correspondingly modified $\tilde m_{\bf kq}$.

From Fig.~2 it is evident that the spin-fluctuation contribution is
dominant over the MF term. When discussing the role of the $J$ term in
the coupling, Eq.~(\ref{mkq}), we note that in the most relevant
region, i.e., along the AFM zone boundary $\tilde m_{\bf
kQ}=2J-4t^\prime \cos^2 k_x$. Thus, for hole doped cuprates,
$t^\prime<0$ and $J$ terms enhance each other in the coupling, and
neglecting $J$ in $\tilde m_{\bf kq}$ reduces $T_c$, although at the
same time relevant $Z_{\bf k}$ is enhanced.

\begin{figure}[htb]
\centering
\epsfig{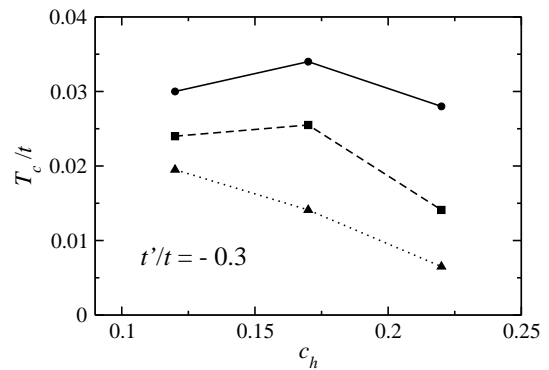}
\caption{$T_c/t$ vs. doping $c_h$ for $t'/t=-0.3$, calculated for
various versions of Eq.~(\ref{del}): a) full result (full line), b)
with neglected MF term (dashed line), and c) in addition to b)
modified $\tilde m_{\bf kq}$ without the $J$ term (dotted line).}
\label{fig2}
\end{figure}

\begin{figure}[htb]
\centering
\epsfig{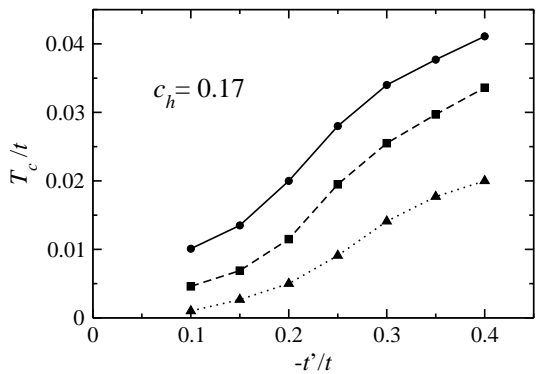}
\caption{$T_c/t$ vs. $-t'/t$ for fixed 'optimum' doping $c_h=0.17$ and
different versions of Eq.~(\ref{del}), as in Fig.~2.}
\label{fig3}
\end{figure}

Finally, in Fig.~3 we present results, as obtained for fixed OP
$c_h=0.17$, but different $t'/t<0$, as relevant for hole-doped
cuprates \cite{pava}. As expected the dependence on $t'$ is
pronounced, since the latter enters directly the coupling $\tilde
m_{\bf kQ}$.  It is instructive to find an approximate BCS-like
formula which simulates our results. The latter involves the
characteristic cut-off energy $\Gamma_{\bf Q}$, while other relevant
quantities are the electron density of states ${\cal N}_0$ and $Z_m$
being the minimum $Z_{\bf k}$ on the FS (in the antinodal
point). Then, we get a reasonable fit to our numerical results with
the expression,
\begin{equation}
T_c \sim 0.5 \Gamma_{\bf Q} ~{\rm e}^{-2/( {\cal N}_0 V_{eff}) },
\label{tc}
\end{equation}
where the effective interaction is given by $V_{eff}= 3 Z_m
(2J-4t^\prime)^2 \chi_{\bf Q}$. Our numerical analysis suggests that
the main $t^\prime$-dependence of $T_c$ originates in the coupling
$\tilde m_{\bf kq}$, not in ${\cal N}_0 Z_m$, while the main
$c_h$-dependence comes from $\chi_{\bf Q}$ and $\Gamma_{\bf Q}$. Then,
Eq.~(\ref{tc}) implies that optimum doping, where $T_c$ is maximum,
increases with $-t'/t$. For parameters used in Fig.~1, e.g.,
$c_{opt}=0.13+0.12(-t'/t)$.

In conclusion, let us comment on the relevance of the present method
and results to cuprates. Our starting point is the model,
Eq.~(\ref{tj}), where strong correlations are explicitly taken into
account via the projected fermionic operators. In this respect the
derivation crucially differs from the analysis of the phenomenological
spin-fermion model \cite{mont}. Nevertheless, in the latter approach
the resulting gap equation, Eq.~(\ref{del}), looks similar but
involves a constant effective coupling. In contrast, our $\tilde
m_{\bf kq}$, Eq.~(\ref{mkq}), is evidently ${\bf k},{\bf q}$-
dependent. In particular, in the most relevant region, i.e., along the
AFM zone boundary, $\tilde m_{\bf kQ}$ depends only on $t^\prime$ and
$J$, but not on $t$. This explains our novel result, i.e., a
pronounced dependence of $T_c$ on $t^\prime$ which is also consistent
with the evidence from different families of cuprates \cite{pava}. One
can give a plausible explanation of this effect. In contrast to NN
hopping $t$, the NNN $t^\prime$ represents the hopping within the same
AFM sublattice, consequently in a double unit cell fermions couple
directly to low-frequency AFM paramagnons, analogous to the case of FM
fluctuations generating superfluidity in $^3$He \cite{legg}.

It is evident from our analysis, that actual values of $T_c$ are quite
sensitive on input parameters and NS properties. Since we employ the
lowest-order mode-coupling approximation in a regime without a small
parameter, one can expect only a qualitatively correct behavior.
Still, calculated $T_c$ are in a reasonable range of values in
cuprates. We also note that rather modest 'optimum' $T_c$ value within
presented spin-fluctuation scenario emerge due to two competing
effects in Eqs.~(\ref{del}),(\ref{tc}): large $\tilde m_{\bf kq}$ and
$\chi_{\bf Q}$ enhance pairing, while at the same time through a
reduced $Z_{\bf k}$ and cutoff $\Gamma_{\bf Q}$ they limit $T_c$.

We do not discuss in detail results for $\Delta_{\bf k}$ in the SC
phase, where we obtain the expected $d_{x^2-y^2}$ form with
$\Delta_{\bf k} \sim \Delta_0(\cos^2 k_x-\cos^2 k_y)/2$, with
$\Delta_0(T=0)\sim \eta T_c$ and $\eta \sim 2.5$. However, we observe
that in the UD regime the effective coherence length $\xi \sim
v_F/\Delta_0(T=0)$ becomes very short. I.e., with $v_F$ taken as the
average velocity over the region $\kappa$ at the antinodal part of the
FS we get $\xi$ ranging from $\xi = 4.4$ in the OD case, to $\xi=1.3$
in the UD example. In the latter case, SC pairs are quite local and
the BCS-like approximation without phase fluctuations,
Eqs.~(\ref{del}),(\ref{tc}), overestimates $T_c$. Starting from this
side, a more local approach would be desirable.

It should also be noted that in the UD regime we are dealing with the
strong coupling SC. Namely, we observe that ${\cal N}_0 V_{eff}$ shows
a pronounced increase at low doping mainly due to large $\chi_{\bf
Q}$.  Then it follows from Eq.~(\ref{tc}) that $T_c$ is limited and
determined by $\Gamma_{\bf Q}$. At the same time, INS experiments
\cite{fong} reveal that in th UD cuprates the resonant peak at $\omega
\sim \omega_r$ takes the dominant part of intensity of ${\bf q} \sim
{\bf Q}$ mode which becomes underdamped possibly even for
$T>T_c$. Thus it is tempting to relate $\Gamma_{\bf Q}$ to $\omega_r$
(for more extensive discussion see Ref.~\cite{sega}) and in the UD
regime to claim $T_c \sim a \omega_r$, indeed observed in cuprates
\cite{fong} with $a \sim 0.26$.  However, additional work is needed to
accommodate properly an underdamped mode in our analysis.

Authors acknowledge the stimulating discussions with N.M. Plakida and
the support of the Ministry of Higher Education and Science of
Slovenia under grant P1-0044.

 \end{document}